\documentstyle[aps,eqsecnum,epsfig]{revtex}

\widetext

\begin{document}
\title{Toward scalable quantum computation with cavity QED systems}
\author{V. Giovannetti$^{1}$, D. Vitali$^{1}$, 
P. Tombesi$^{1}$, and A. Ekert$^{2}$}
\address{$^{1}$Dipartimento di Matematica e Fisica, Universit\`a di
Camerino, \\
INFM, Unit\`a di Camerino, via Madonna delle Carceri 62032, Camerino, Italy \\
$^{2}$Centre for Quantum Computation, Clarendon Laboratory,
Parks Road, Oxford OX1 3PU, UK}
\date{\today}
\maketitle

\begin{abstract}
We propose a scheme for quantum computing using high-Q cavities
in which the qubits are represented by single
cavity modes restricted in the space spanned by the two lowest 
Fock states. We show
that single qubit 
operations and universal multiple qubit gates can be 
implemented using atoms sequentially crossing the
cavities.
\end{abstract}

\pacs{42.50.Lc, 03.65.-w}

\section{Introduction}

In the last years, numerous physical systems have been proposed
as possible candidates for the implementation of a quantum computer.
The desirable conditions which have to be satisfied are
a reliable and
easy way to prepare and detect the quantum states of the qubits, 
the possibility to engineer highly entangled states,
the scalability to large number of qubits and a very low decoherence rate
\cite{bendivi}.
Up to now, experimental implementations have involved 
linear ion traps \cite{cizo}, liquid state NMR 
\cite{nmr}, and cavity QED systems \cite{turch,qpg}. 
In the ion trap case, only the controlled-not
(C-NOT) gate between two internal states and the vibrational level of a single
ion has been realized \cite{cnot}, and quantum gates involving
two or more ions
have not yet realized experimentally. A promising step in this direction is the
very recent generation of an entangled state of four ions, even if only with
$57 \%$ fidelity \cite{winnatu}.
The status of liquid-state NMR quantum computing is still debated \cite{caves},
but the fact that the signal strength becomes exponentially small with the
number of qubits makes this proposal certainly not
scalable to more than about ten qubits. This explains why the 
research of new physical implementations of a quantum computer is so active
(see \cite{divin} and references therein).
Here, elaborating on the suggestions of Ref.~\cite{Eke},
we propose to use the Fock's states  
$| 0 \rangle$ and $| 1 \rangle$ of a high-$Q$ cavity mode
as the two logical states of a qubit. 
A quantum register of $N$ qubits is therefore a collection of 
$N$ identical cavities in which the state of an appropriately chosen
cavity mode is within the space spanned by the vacuum and the one photon state.
The register transformations are achieved sending off-resonant two-level 
atoms through the cavities and making them mutually interacting
by means of suitable classical fields. With this respect, the present proposal 
is similar to that of Refs.~\cite{ticho,Domo}; 
the important difference is that, in these papers, the 
logical qubits are represented by two circular Rydberg levels of the atoms.
In our proposal, the role of atoms and cavity modes are exchanged.
In this way, the present scheme becomes scalable in principle.
In practice, its scalability can be limited 
by the spontaneous emission from
the Rydberg levels or by other technical limitations, 
but the present proposal has the advantage
that the needed technology is essentially already available to realize
some proof-of-principle demonstrations of quantum computation 
with few qubits.
In fact, in the present paper we shall specialize 
to the case of microwave cavities,
for which a high level of quantum state control and engineering
has been already experimentally shown \cite{qpg,haros}. This is the
reason why in our explicit
calculations we shall consider microwave cavities  
operating in low-order modes with angular frequency $\omega$ in the 
$10-100$ GHz range, and Rydberg atoms for which high values of the   
coupling constant (of the order of $10^{5}$ $s^{-1}$) are 
possible. It is clear however that, in principle, the method can be applied
to optical cavities too, in which one can have a miniaturization of the 
scheme and therefore a faster operation.

Some preliminary results on the possibilities offered by the present 
cavity QED scheme have been shown in \cite{jmo}, where some 
implementations of the C-NOT gate between two cavity modes have 
been presented. In order to give a clear and exhaustive 
description, here we shall review the results
of \cite{jmo}, which will be extended and generalized in the present 
paper.

The outline of the paper is as follows. In Sec.~II we review the basic
properties of the considered cavity QED system. In Sec.~III we show
how to implement the universal C-NOT gate between two cavities, while in
Sec.~IV we shall discuss a different scheme for the implementation
of universal two-qubit gates, using an arrangement based on that 
adopted in the experiment of Ref.~\cite{haros}. In Sec.~V
we show how single qubit operations can be realized, while
Sec.~VI is devoted to the implementation of useful many-qubit
universal quantum gates as the Toffoli gate and the encoding
and decoding network for
quantum error correction schemes. Section VII is
for concluding remarks, while the appendix shows the explicit
implementation of the Deutsch problem \cite{Deut1,Deut2}.

\section{The system}
 
The interaction of a two-level atom quasi-resonant with
a high-Q cavity mode is well described by 
the time dependent Hamiltonian \cite{Haro}:
\begin{equation}
{\cal H}(t)= \frac{\hbar \omega}{2} \big[ b^{\dagger}  b + b b^{\dagger} \big] 
+\frac{\hbar \omega_{eg}}{2} \big[ | e \rangle \langle e | - | g \rangle 
\langle g | \big] + \hbar \Omega(t) \big[ | e \rangle \langle g | b  + | g \rangle 
\langle e | b^{\dagger} \big],
\label{uno}
\end{equation}
in which $b$ and $\omega$ are the annihilation operator and the 
angular frequency of the cavity mode respectively; $| e \rangle $
and $| g \rangle$  are the excited and lower circular Rydberg state and 
$\hbar \omega_{eg}$ is their energy difference. Finally 
$\Omega(t)$ is the atom-field interaction Rabi frequency which is 
time dependent because of the atomic motion through the cavity. In 
particular for a Fabry-P\'{e}rot-type cavity, with Gaussian transverse 
beam profile, we can assume the following continuous variation
\begin{equation}
\Omega(t)=\Omega_{0} e^{- (\frac{t}{\tau})^{2}},
\label{due}
\end{equation}
where $2 \tau$ is the atomic transit time, which 
depends of course on the inverse of the atomic velocity. 
For $|t| \gg \tau$, i.e. when the atom is outside 
the cavity, the 
energy eigenvectors of the system are $|g \rangle 
\otimes |n\rangle \equiv |g,n\rangle$ and  $|e,n\rangle$, with $|n\rangle $ the generic
Fock state of the cavity mode. Apart for the ground state $| g,0 
\rangle$ which remains unchanged, in the presence of the time-dependent 
interaction, these terms are coupled by photon emission or 
absorption, and the istantaneous energy eigenstates at fixed time $t$ are
the dressed states
\begin{equation}
| {\cal V}^{(n)}_{\pm}(t) \rangle = \frac{ (\delta /2  \pm \sqrt{(\delta 
/2)^{2} + \Omega^{2}(t)(n+1)}) |e,n\rangle + \Omega(t)\sqrt{n+1} 
|g,n+1\rangle }{\sqrt{\delta^{2}/2 +2\Omega^{2}(t)(n+1) \pm \delta \sqrt{(\delta 
/2)^{2} + \Omega^{2}(t)(n+1)}}}
\label{tre}
\end{equation}
with eigenvalues 
\begin{equation}
 E^{(n)}_{\pm}(t) = \hbar \omega (n+1) \pm \hbar \sqrt{(\delta 
/2)^{2} + \Omega^{2}(t)(n+1)},
\label{quattro}
\end{equation}
where $\delta=\omega_{eg}-\omega$ is the atom-cavity 
detuning. Fig.~1 qualitatively shows the time dependence of the dressed 
levels of Eq.~(\ref{quattro}) in the case $\delta>0$. 
Now, if the atom velocity is slow enough, and the system for 
$t\ll -\tau$ is prepared in a generic energy eigenstate, then in its
time evolution it will adiabatically follow 
this eigenstate, with 
negligible transitions toward other states \cite{messiah}. 
The exact adiabatic condition 
can be obtained writing the Schr\"{o}dinger equation in the 
basis of the vectors $| {\cal V}^{(n)}_{\pm}(t) \rangle$, and then
neglecting terms coupling the dressed states. The resulting condition 
is
\begin{equation}
\frac{\dot{\Omega}(t) \,  \delta \, \sqrt{n}}{4 \big[ (\delta 
/2)^{2} + \Omega^{2}(t)n\big]^{3/2}} \ll 1,
\label{cinque}
\end{equation}
which, in the limit $ \Omega(t) \sqrt{n}/ \delta \ll 1$ becomes 
equal to that given in Ref. \cite{Haro}. The general
adiabatic condition (\ref{cinque}) shows in particular that
adiabaticity can be obtained even when
$ \Omega(t) \sqrt{n}/ \delta \simeq 1$, provided that 
$\dot{\Omega}(t)$ is sufficiently small. In the following we shall 
always work in this adiabatic regime.

\section{The C-NOT gate}

Domokos {\em et al.}  have shown in Ref.~\cite{Domo} that, using 
induced transitions between the dressed states, it is possible to 
implement a C-NOT gate in which a cavity containing at most
one photon is the control qubit and the atom is the target qubit. 
This idea is the starting point for the implementation of the C-NOT
between two cavities we propose here.
Ref.~\cite{Domo} considers an atom entering adiabatically the cavity so 
that the joint atom-cavity state is
\begin{equation}
 c_{1} |g,0\rangle + c_{2}|g,1\rangle + c_{3}|e,0\rangle + 
 c_{4}|e,1\rangle.
 \label{sei}
 \end{equation}
When the atom is just inside the cavity, 
 a classical field $S$ of frequency 
 $\omega_{S}$ equal to the energy difference between the dressed states 
  $| {\cal V}^{(1)}_{+}(t=0) \rangle$ 
 (originating from $|e,1\rangle$) and $| {\cal V}^{(0)}_{-}(t=0) \rangle$
 (originating from $|g,1\rangle$) 
  is switched 
on for a time interval $2 \tau_{S}$, so that the following driving Hamiltonian
is added to ${\cal H}(t)$ of Eq. (\ref{uno})
  \begin{equation}
  {\cal H}_{S}(t)= - \hbar \Xi_{0} \cos ( \omega_{S}t + \varphi_{S})
   e^{- (\frac{t}{\tau_{S}})^{2}} \big[ | e \rangle \langle g | 
   + | g \rangle\langle e | \big],
  \label{sette}
\end{equation}
where $\varphi_{S}$ is the phase of the classical 
field $S$ and 
$\Xi_{0}$ is the coupling costant, which depends on the dipole 
moment of the transition $ e  \leftrightarrow g$ and on the intensity of $S$.
Appropriately choosing the value of $\tau_{S}$, 
 it is now possible to selectively couple
 $S$ with these dressed 
 states, leaving the other 
 components of the vector state essentially unperturbed. Moreover, with 
 a suitable choice of the intensity $S$, it is possible to apply 
a Rabi $\pi$ pulse between the two states.
 In this way, when the atom exits the cavity,
 the resulting state vector in the interaction picture, 
 apart for some phase terms,
 is given by Eq.~(\ref{sei}), but with $|e,1\rangle$ and 
 $|g,1\rangle$ exchanged. 
In this way one has realized a C-NOT gate in which, when the cavity 
(the control
 qubit) has one photon, the atom undergoes a NOT operation, while 
 when the cavity contains no photons, the atomic state remains unchanged.
 We shall refer to this gate as the 
 C-NOT(cavity $\rightarrow$ atom).
 
In a similar manner, we can build also a C-NOT gate in which
the roles of atom and cavity are exchanged. Let us assume in fact
to tune the frequency $\omega_{S}$ to the transition between the 
dressed state $| {\cal V}^{(0)}_{-}(t=0) \rangle$ and the state
$|g,0\rangle$, and apply again a $\pi$ pulse inside the cavity
as before.
Now, when the atom leaves the cavity, 
the terms $|g,0\rangle$ and $|g,1\rangle$ in the vector state of the system
are mutually exchanged with respect to the initial condition
 Eq.~(\ref{sei}). The $|e,0\rangle$ and $|e,1\rangle$ 
components are instead not affected by the interaction with the
classical source $S$.
This means having realized a C-NOT gate in which, when the atom is in the 
  ground state, the cavity states $|0 \rangle$ and $|1 \rangle$ flip,
  while nothing happens to the cavity
  state for the atom in the excited state.
In analogy with the previous case, we refer to this new gate
as C-NOT(atom $\rightarrow$ cavity).
  
It is important 
to note that, differently from the C-NOT(cavity $\rightarrow$ atom) case,
in the C-NOT(atom $\rightarrow$ cavity) gate 
the Rabi transition between the original
states (i.e.  $|g,0\rangle$ and $|g,1\rangle$) 
of the dressed states involved
is forbidden by selection rules. Nevertheless, this coupling is 
realizable when the atom is in the cavity because the vector
$| {\cal V}^{(0)}_{-}(t=0) \rangle$ has also 
a $|e,0\rangle$ component. 
However, since this dependence
is mediated by a coefficient which decreases 
with $\Omega_{0}/ \delta$ (see Eq. (\ref{tre})),
we have to choose a not too small value of this 
parameter in order to have a significative coupling constant. 
In particular, in our calculations
we have chosen $\Omega_{0}/ \delta \sim 10^{-1}$.
With such values for $\delta$, it is also possible to have a sufficient 
frequency separation between the transitions we are interested in and
all the other ones. Actually, it is sufficient to set the duration 
of the classical pulse $2 \tau_{S}$ of the order of $20 \mu s$
to discriminate all the parasitic transitions and optimal results 
for the resulting quantum operation are 
achieved for $\Omega_{0}=420$ kHz, $\delta= 0.18 \,\Omega_{0}$,
$\tau \sim 100 \mu s$ and with $\Xi_{0}=240$ kHz,
$\tau_{S}=14 \mu s$ for the C-NOT(cavity $\rightarrow$ atom),
$\Xi_{0}=141.5$ kHz, $\tau_{S}=19 \mu s$ for the C-NOT(atom $\rightarrow$ cavity).
In Fig.~2 we show the time evolution of the dressed states 
populations for the C-NOT(atom $\rightarrow$ cavity) for the above choice
of parameter values.
  
For quantum information processing one needs to control 
not only the level populations, 
but also the relative phases. 
In general, during the adiabatic evolution, different dynamical phases
for the different
components of the vector state are generated. However, it is always
possible to correct these phases
by an appropriate choice of the field phase $\varphi_{S}$ 
and eventually acting outside the cavity on the atom with suitable
Stark electric fields.
  
We now have all the elements to realize the C-NOT gate between two 
distinct but identical cavities, $A$ and $B$, with the first one
acting as the control qubit and the second one as the target qubit. 
The apparatus is sketched in Fig.~3 and it is 
essentially a physical realization of the logical network shown in
Fig.~4. Suppose that the initial states 
of the two cavities are respectively $
|\phi\rangle_{A}=\alpha_{A}|0\rangle_{A} + \beta_{A}|1 
\rangle_{A}$ and 
$ |\psi\rangle_{B}=\alpha_{B}|0\rangle_{B} + \beta_{B}|1 
\rangle_{B}$. 
{\em i)} A first atom, $a_{1}$, prepared in the 
ground state $|g \rangle$, enters cavity $A$, where it
undergoes the C-NOT(cavity $\rightarrow$ atom) transformation realized
with the classical field source $S_{A}$, and described above.
{\em ii)} Then $a_1$ leaves $A$ and enters cavity $B$: here 
the classical field $S_{B}$ is switched on in 
order to obtain a C-NOT(atom $\rightarrow$ cavity) transformation. 
In the interaction picture and neglecting all 
the parasitic but controllable phase 
terms, the state of the total system at this stage is then:   
\begin{equation}
  	\alpha_{A}| 0 \rangle_{A}\otimes \overline {| \psi \rangle}_{B} \otimes |g \rangle+
  	\beta_{A}| 1 \rangle_{A}\otimes | \psi\rangle_{B} \otimes |e \rangle,
  	\label{nove1}
  \end{equation}
  where $\overline {| \psi \rangle}_{B}$ is  the 
  NOT-conjugate vector of $| \psi\rangle_{B}$, that is, 
  $\beta_{B}|0\rangle_{B} + \alpha_{B}|1 \rangle_{B}$.
  {\em iii)} and {\em iii)} The atom enters again $A$, where it 
  undergoes the C-NOT(cavity $\rightarrow$ atom) transformation, 
so that the state of the system becomes
  \begin{equation}
  	\alpha_{A}| 0 \rangle_{A} \otimes \overline {| \psi \rangle}_{B} \otimes |g \rangle+
  	\beta_{A}| 1 \rangle_{A}\otimes | \psi\rangle_{B} \otimes |g 
  	\rangle = 
  	\left\{ \alpha_{A}| 0 \rangle_{A} \otimes \overline {| \psi \rangle}_{B} +
  	\beta_{A}| 1 \rangle_{A}\otimes | \psi\rangle_{B} \right\} \otimes |g \rangle.
  	\label{novebis}
  \end{equation}
  In terms of the notations given in the previous section we shall refer 
  to this gate as the C-NOT-INV($A \rightarrow B$)
gate, in order to underline that $B$ undergoes to a 
  NOT transformation when $A$ is in the $|0 \rangle_{A}$ state,
  while nothing happens when
  $A$ is in the $|1 \rangle_{A}$ state. This fact is illustrated in 
  Fig.~4, where in the equivalent gate {\bf (II)} there is a NOT
  transformation acting on $B$.
  
The practical realization of step {\em iii)}, 
i.e. the return of the atom in the first cavity, is 
  actually more complicated than what it looks from Figs.~3 and 4. 
The inversion of the motion of atom $a_1$, could be realized in
principle with an atomic fountain configuration. However this implies 
having free-fall velocities, which are too slow in order to have
the necessary interaction times within the cavities. For this reason  
we propose to transfer the quantum information from this atom 
onto a second one of the same type, but travelling in the opposite 
direction. With this respect, the scheme adopts the 
``quantum memory'' scheme experimentally verified in Ref.~\cite{memory}.
This quantum information transfer is implemented introducing 
  a third cavity, the auxiliary cavity $M$ of Fig.~3, 
  which, differently from $A$ and $B$,
is resonant with the $e\rightarrow g$ transition. If $M$ is 
  prepared in the vacuum state $|0 \rangle_{M}$, and the transit time of 
  $a_{1}$ is appropriately chosen, then the atomic state component $|e \rangle$ 
  releases one photon in $M$ through a resonant $\pi$ Rabi oscillation. 
After that, the 
  state of the total system (the three cavities and $a_{1}$), using 
the same notations of Eq.~(\ref{nove1}), will be
  \begin{equation}
  	\alpha_{A}| 0 \rangle_{A}\otimes \overline {| \psi 
  	\rangle}_{B} \otimes | g \rangle \otimes| 0 \rangle_{M}+
  	\beta_{A}| 1 \rangle_{A}\otimes | \psi\rangle_{B}
  	\otimes |g \rangle \otimes| 1 \rangle_{M}.
  	\label{dieci1}
  \end{equation}
Notice that the entanglement of $a_{1}$ with 
  $A$ and $B$ is now transferred to the auxiliary 
  cavity $M$: the state of the atom $a_{1}$ is factorized
and it can be neglected from now on. At this 
stage, a second 
  atom $a_{2}$ is prepared in the ground state $|g \rangle$ 
  and injected into the apparatus with the same absolute value of 
  velocity of $a_{1}$, but with the opposite direction. 
  Entering $M$, it absorbes the photon left by the 
  first atom through a similar $\pi$ Rabi oscillation,
  and the entanglement with the cavities $A$ and $B$
  is transferred from $M$ to $a_{2}$:
  \begin{equation}
  	\alpha_{A}| 0 \rangle_{A}\otimes \overline {| \psi 
  	\rangle}_{B} \otimes | g \rangle \otimes| 0 \rangle_{M}+
  	\beta_{A}| 1 \rangle_{A}\otimes | \psi\rangle_{B}\otimes |e 
  	\rangle \otimes| 0 \rangle_{M}.
  	\label{1uno}
  \end{equation}
At this stage also the state of the cavity $M$ is factorized and therefore
  the vector state (\ref{1uno}) is quantum logically 
  equivalent to that of Eq. (\ref{nove1}). In practice,  
  the apparatus described here is essentially an 
  {\em atomic mirror}, which permits us to {\em reflect back} the 
  atomic state. Finally let us observe that $a_{2}$ has to cross 
cavity $B$ without interacting with it,
  before it could reach the cavity $A$. 
  This result is achieved simply by switching off the classical source $S_{B}$; 
  the adiabatic regime and the 
  off-resonance condition prevents that, apart for the dynamical  
  phase factors, the state could change 
  during the transit of $a_{2}$ within $B$. The action of the 
  C-NOT gate has been explicitely described for factorized state only 
  just to simplify the presentation. It is clear that all the steps 
  can be repeated for a generic entangled state of the two cavities.
   
  Assuming the optimal values for the system parameters written above, 
  we have solved numerically the time evolution 
  of the total system. We describe the resulting
  effective C-NOT gate in the form of
  a matrix written in the basis of the Fock states
  of the two cavities, $\left\{ | 0, 0 \rangle; 
  | 0, 1 \rangle; | 1, 0 \rangle; | 1,1 \rangle  \right\}$
  where  $| n, m \rangle=| n \rangle_{A}\otimes | m \rangle_{B}$:
  The matrix has been ``cleaned up'' of the spurious phase factors which 
may appear during the evolution, and which, using the phase of the 
  classical field $S_{B}$ and also appropriate Stark shift
  electrical fields, can always be suitably adjusted.  
 Within a $0.1\%$ error, the optimized C-NOT matrix
 has the form
  \begin{equation}
 \left( 
   \begin{array}{ccccccc}
    0 & & e^{-i \lambda} & & 0 & & 0 \\ \\
    e^{ i \lambda} & & 0 & & 0 & & 0 \\ \\
    0 & & 0 & & e^{ i \lambda} & & 0 \\ \\
    0 & & 0 & & 0 &  & e^{-i \lambda}
   \end{array}
   \right) \;,
 \label{1due}
 \end{equation} 
where the non trivial phase $\lambda = 0.07 $. 
The overall transformation takes place in
  a time of the order of one millisecond, which has to be compared
  with the typical decoherence timescales, that is, the atomic
  radiative lifetimes and the cavity relaxation times.
For circular Rydberg atoms with $n\simeq 50$, 
the atomic radiative lifetime is of the order of 
  $30 ms$ and therefore it does not represent a serious
  problem. The cavity damping times currently
  realized for microwaves have instead the same order of magnitude
  (some millisecond). However relaxation times
of the order of $10 ms$ will be hopefully achieved in the near future,
and in this case, one would have a perfectly working C-NOT 
gate between two cavities.
It is clear therefore that, for the present implementation of quantum 
information processing, the main source of decoherence in the microwave domain
is just the cavity leakage. If optical cavities are instead considered,
also atomic spontaneous emission may represent an important source 
of decoherence.

The matrix of Eq.~(\ref{1due}) is not a pure C-NOT-INV gate, 
even if it is still
  an universal two-qubit gate \cite{Eke1};
in particular it can be transformed into a standard C-NOT gate by adding
a single qubit operation on $B$, similar to those we shall present in the following 
  section.
  Moreover, it is also possible to
  implement the C-NOT($A \rightarrow B$) gate (i.e. the
  one in which the vector component $|1 \rangle_{A}$ causes the
  NOT transformation on $B$, so that the role of the target and control qubit
  is exchanged), simply 
  preparing the first atom entering the apparatus in
  $| e \rangle$ rather than in $|g \rangle$, and then proceeding
  with the same identical steps of the C-NOT-INV case.
  
  The above scheme assumes the possibility of injecting in the 
  apparatus atomic pulses 
  with exactly one atom: this is not experimentally achieved up to now.
  However, as shown in Ref.~\cite{Mauro}, a 
  control of the atom number could be achieved by a modification of 
  the Rydberg atoms preparation technique. 
The idea is to use an ``atom counter'' before the
circular Rydberg state preparation, so that the preparation of the
state $|e\rangle $ (which can have an efficiency near to $100\%$)
is applied only when one is sure to have exactly one atom.
The atom counter is realized by driving a strong
transition and measuring the fluorescence, whose intensity
wiil be proportional to the number of atoms. 
  When the beam section contains zero, two or more atoms, it is discarded:
  the system waits then for a time of the order of a few microseconds
  to twenty microseconds (depending upon the atomic velocity and the
  precise length of the atomic beam section) until a fresh section
  of the beam comes in the laser beam driving the fluorescence. 
In this way, instead of preparing a random number at a given time, one thus 
  prepares with a high probability a single Rydberg atom after a
  random delay.
  The average delay is minimal when the probability to
  have exactly one atom is maximized.
  With a Poissonian statistics, the optimal mean number of atoms is
  $1$. The average random delay could be
  of the order of $25 \mu s$ in realistic experimental condition.
  This is short enough at the scale of the cavity field lifetime
  to play no major role in the proposed scheme.
  
\section{Quantum phase gate}

In the preceding section we have seen how to implement a universal
two-qubit gate, the C-NOT gate, between two cavity modes, using 
induced transitions between dressed states. However, it is possible to
realize another universal two-qubit gate, the quantum phase gate (QPG)
\cite{turch,qpg2},
between the two cavities, slightly elaborating on the quantum phase gate
operating on qubits carried by the Rydberg atom and the two lowest 
Fock states of a cavity mode, recently demonstrated experimentally 
\cite{qpg}. Of course, since both the C-NOT and the QPG are universal 
quantum gates, it is always possible to implement one of them, by simply 
supplementing the other one with appropriate one-qubit operations.
However, the interesting experimental result of Ref.~\cite{qpg}
suggests an alternative physical implementation of quantum logic 
operations between cavity qubits, which does not involve induced 
transitions between dressed states, and extends the scheme of 
\cite{qpg} to a directly scalable model.

The QPG transformation reads
\begin{equation}
|a,b\rangle \rightarrow \exp\left(i\phi 
\delta_{a,1}\delta_{b,1}\right) |a,b\rangle ,
\label{qpgeq}
\end{equation}
where $|a\rangle $ and $b\rangle $ describe the basis states 
$|0\rangle $ and $|1\rangle $ of two generic qubits.
This means that the QPG leaves the initial state unchanged unless 
when both qubits are in state $|1\rangle$. In Ref.~\cite{qpg}, the 
QPG of Eq.~(\ref{qpgeq}) did not involve levels $g$ and $e$, but
$i$ and $g$, where $i$ is a lower circular Rydberg level, which is 
uncoupled with the high-Q cavity. In this way, the gate of 
Eq.~(\ref{qpgeq}) in the case $\phi =\pi$ can be realized by setting
the atomic transition $g\rightarrow e$ perfectly at resonance with 
the relevant cavity mode (by appropriately Stark-shifting the atomic 
levels inside the cavity), and by selecting the atomic velocity
so that the atom undergoes a complete $2\pi$ Rabi pulse when crossing 
the cavity. In fact, at resonance, such a pulse transforms the state
$|g,1\rangle $ into $e^{i\pi}|g,1\rangle $, while nothing happens if
the atoms is in $i$ or the cavity is in the vacuum state. In 
\cite{qpg}, the possibility to tune the phase $\phi$ over a large 
range, by slightly detuning the cavity mode from the $g \rightarrow e$
transition, has been shown, but we shall not consider this possibility 
here.

We now show that this atom-cavity QPG can be used to realize a QPG 
between two cavity modes, by considering an arrangement very similar 
to that of the C-NOT gate shown in Fig.~3. The cavities $A$ and $B$ are
again the two qubits, while $M$ is again the auxiliary cavity needed to
``reflect'' the atom and disentangle it. The two classical 
sources inside the cavities $S_{A}$ and $S_{B}$ are no more needed, while we
consider the possibility to apply Stark shift electric fields inside 
the cavities, in order to tune the $g \rightarrow e$ transition in and  
out of resonance from the cavity mode. The scheme of the QPG 
implementation is shown in Fig.~5 and involves only two atom crossings, 
as in the C-NOT gate of the preceding section, and three $\pi/2$ pulses 
between the $i$ and $g$ levels (the Hadamard gates $H$ of Fig.~5), 
which can be realized with resonant classical microwave sources applied
between the high-Q cavities. 

Let us assume a generic state of the two cavity qubits 
\begin{equation}
|\psi\rangle =
a_{0}|00\rangle +a_{1}|01\rangle + a_{2}|10\rangle + a_{3}|11\rangle 
\label{psize}
\end{equation}
and that a first atom, initially prepared in state $i$, is subject to
a $\pi/2$ pulse, so to enter cavity $A$ in state $\left(|i\rangle +
|g \rangle \right)/\sqrt{2}$. The cavity mode is perfectly resonant
with the $g \rightarrow e$ transition and the atom velocity is selected
so that the atom undergoes a $2\pi$ Rabi pulse if it is in state $g$
and the cavity contains one photon (the QPG of Ref.~\cite{qpg}). 
The resulting state at the exit 
of cavity $A$ is
\begin{equation}
\frac{|i\rangle \otimes |\psi\rangle}{\sqrt{2}} +
\frac{|g\rangle}{\sqrt{2}} \otimes \left(
a_{0}|00\rangle +a_{1}|01\rangle - a_{2}|10\rangle - a_{3}|11\rangle
\right).
\label{psi2}
\end{equation} 
Then the atom undergoes another resonant $\pi/2$ pulse on the
$i \rightarrow g$ transition and the state of the system becomes
\begin{equation}
|i\rangle \otimes \left(a_{0}|00\rangle +a_{1}|01\rangle\right)
+|g\rangle \otimes \left(a_{2}|10\rangle + a_{3}|11\rangle
\right).
\label{psi2bis}
\end{equation} 
Then the atom crosses cavity $B$, where it is subjected 
again to the atom-cavity QPG as in cavity $A$, so that
the state of the system becomes
\begin{equation}
|i\rangle \otimes \left(a_{0}|00\rangle +a_{1}|01\rangle\right)
+|g\rangle \otimes \left(a_{2}|10\rangle - a_{3}|11\rangle
\right).
\label{psi3}
\end{equation} 
At this point, as in the C-NOT case of the preceding section, in order
to realize the final transformation and disentangle the atom from the 
cavities, one has to ``reflect'' it. This is again achieved with 
the ``atomic mirror'' scheme, i.e., using the auxiliary cavity $M$ of 
Fig.~5, that acts as a quantum memory and is able to transfer the 
entanglement from the first atom to a second atom, which is crossing 
the apparatus with the same absolute velocity but in the opposite 
direction (see Eqs.~(\ref{dieci1}) and (\ref{1uno})). 

The second atom is then subjected to a $\pi/2$ pulse before entering 
the cavities and the state of Eq.~({\ref{psi3}) becomes
\begin{equation}
\frac{|i\rangle}{\sqrt{2}} \otimes \left(
a_{0}|00\rangle +a_{1}|01\rangle + a_{2}|10\rangle - a_{3}|11\rangle
\right) +
\frac{|g\rangle}{\sqrt{2}} \otimes \left(
a_{0}|00\rangle +a_{1}|01\rangle - a_{2}|10\rangle + a_{3}|11\rangle
\right).
\label{psi4}
\end{equation}
The last step is the QPG between the atom and cavity $A$, i.e., the 
atom has to cross cavity $B$ undisturbed (this is achieved by strongly 
detuning the $g \rightarrow e$ transition with a Stark shift field)
and then has to undergo another full Rabi cycle in cavity $A$.
The final state is
\begin{equation}
\frac{|i\rangle +|g \rangle }{\sqrt{2}} \otimes \left(
a_{0}|00\rangle +a_{1}|01\rangle + a_{2}|10\rangle - a_{3}|11\rangle
\right) ,
\label{psi5}
\end{equation}
which is desired result, corresponding to a QPG between cavities $A$ 
and $B$ with conditional phase shift $\phi = \pi$, 
and with a disentangled atom.

\section{One qubit operations}
 
One qubit operations are straightforward to implement on qubits 
represented by two internal atomic states because it amounts to apply
suitable Rabi pulses. This task is less trivial for bosonic degrees of 
freedom as our cavity modes, because the two lowest Fock states for 
example are coupled to the more excited ones.
The most practical solution is to implement one-qubit operations 
on the two lowest Fock states
sending again atoms through the cavity. To be more specific,
one has to send an atom prepared in the ground state $|g\rangle$ 
through the cavity, with the classical field $S$ tuned at the 
frequency corresponding to the transition between the states 
$| {\cal V}^{(0)}_{-}(t=0) \rangle$ and $|g,0\rangle$. 
If one sets the time duration and the intensity of the classical
source $S$ as in the case of the C-NOT(cavity $\rightarrow$ atom), i.e.
such to realize a $\pi$ pulse between the selected levels, one 
implements a ``not-phase'' gate, which, in 
the canonical basis  $\left\{ | 0 \rangle ,| 1 \rangle \right\}$, 
 is described by the following matrix
  \begin{equation}
 N(\theta)=\left( 
   \begin{array}[c]{ccc}
    0 & & e^{-i \theta}\\ \\
    e^{ i \theta} & & 0
   \end{array}
   \right),
 \label{1tre}
 \end{equation}
 where $\theta$ depends linearly on the phase $\varphi_{S}$ of the
 classical field of Eq.~(\ref{sette}) and it is therefore 
 easily controllable (this scheme is 
simply a part of the C-NOT-INV($A \rightarrow B$)
gate presented before). If otherwise the atom inside the cavity undergoes 
a $\pi/2$ instead of a $\pi$ pulse,
one realizes the ``Hadamard-phase'' gate
  \begin{equation}
 H(\theta^{\prime})=\frac{1}{\sqrt{2}} \left( 
   \begin{array}[c]{ccc}
    1 & & e^{-i \theta^{\prime}}\\ \\
    e^{i \theta^{\prime}} & & -1
   \end{array}
   \right),
 \label{1quattro}
 \end{equation}
where also $\theta^{\prime}$ depends linearly on the
classical field phase $\varphi_{S}$ and is therefore controllable.
$N(\theta)$ and $H(\theta^{\prime})$ 
can be used to build the more general 
one-qubit operation and therefore, together
with C-NOT-INV($A \rightarrow B$), form
an universal set of gates. 

Note that the not-phase gate can be used also for an alternative
realization of the C-NOT-INV($A \rightarrow B$)
gate between two cavities. In the scheme described in the preceding
section, one needs the auxiliary cavity M and the second atom crossing
in the opposite direction in order to disentangle the first atom from 
the cavities. One could simplify this last stage (step {\it iii)} and 
{\it iv} of the preceding section) by applying an exact $\pi/2$ pulse
when the atom has just left the second cavity and the state
 of the system is that of 
Eq.~(\ref{nove1}). The total state becomes
 \begin{eqnarray}
   \lefteqn{\alpha_{A}| 0 \rangle_{A}\otimes \overline {| \psi \rangle}_{B} \otimes 
    \frac{|g \rangle+|e \rangle}{\sqrt{2}}+
  	\beta_{A}| 1 \rangle_{A}\otimes | \psi\rangle_{B} \otimes  
  	\frac{|g \rangle-|e \rangle}{\sqrt{2}} = } \\ \label{1cinque}
  	& & \left\{ \alpha_{A}| 0 \rangle_{A} \otimes \overline {| \psi 
  	\rangle}_{B} + \beta_{A}| 1 \rangle_{A}\otimes | \psi\rangle_{B} 
  	\right\} 
  	\otimes |g \rangle/\sqrt{2}+
  	\left\{\alpha_{A}| 0 \rangle_{A} \otimes \overline {| \psi 
  	\rangle}_{B} - \beta_{A}| 1 \rangle_{A}\otimes |  
  	\psi\rangle_{B}\right\}
  	\otimes |e \rangle/\sqrt{2}. \nonumber
  \end{eqnarray}
If now the atom is detected by a state-sensitive detector and
the $|g \rangle$ state is detected, the two cavities are projected on
$\alpha_{A}| 0 \rangle_{A} \otimes \overline {| \psi 
  	\rangle}_{B} + \beta_{A}| 1 \rangle_{A}\otimes | \psi\rangle_{B}$ 
and one has implemented just the desired C-NOT-INV gate. 
On the contrary, if the atom is found in the excited state 
$|e \rangle$, the state of $A$ and $B$ becomes 
   $\alpha_{A}| 0 \rangle_{A} \otimes \overline {| \psi 
  	\rangle}_{B} - \beta_{A}| 1 \rangle_{A}\otimes | \psi\rangle_{B}$
and the C-NOT-INV gate is obtained once that cavity $A$ is subject 
to a $\pi$-phase shift, which can be realized 
by means of two not-phase gates $N(\theta)$, the first with 
  	$\theta=\pi/2$ and the second with $\theta=0$.
In this way the atom is disentangled by the measurement. 
However, the practical application of this scheme is 
seriously limited by the quantum efficiency of atomic detectors, 
which is usually far from $100 \%$.

\section{Many-qubit gates}

\subsection{The Toffoli gate}

We have shown how to implement a set of universal quantum gates
with the proposed cavity QED scheme.
Therefore, in principle, the most general quantum operation involving
$n$ qubits can be realized in terms of the one and two-qubit 
operations described above. This decomposition however implies 
a degree of network complexity, and a number of resources and steps 
which is rapidly increasing with the number of qubits $n$.
One of the main advantages of the present proposal is that it is
particularly suited for the efficient implementation of 
many-qubit quantum gates, which, in many cases, can be realized with the
same number of steps of the two-qubit C-NOT gate of Section III.

A particularly clear example of the possibilities of the proposed 
scheme is provided by the Toffoli gate 
\cite{toffoli}
\begin{equation}
|x\rangle_{A}|y\rangle_{B}|z\rangle_{C} \rightarrow | x 
\rangle_{A} |y\rangle_{B} \left | \Big[ z + (x \wedge y)
\Big]_{mod2} \right\rangle_{C},
\label{ok1}
\end{equation}
in which the target qubit $C$ is controlled by the first two, $A$ 
and $B$. 
The effect of the Toffoli gate on the generic three qubit state
\begin{equation}
|\Psi_{0}\rangle = \alpha_{1}|000\rangle +\alpha_{2}|001\rangle +\alpha_{3}|010\rangle 
+\alpha_{4}|011\rangle +
\alpha_{5}|100\rangle +\alpha_{6}|101\rangle +\alpha_{7}|110\rangle 
+\alpha_{8}|111\rangle
\label{ok2}
\end{equation}
($|n,m,l \rangle = |n\rangle_{A} \otimes |m\rangle_{B} \otimes 
|l\rangle_{C}$ are the tensor product of the cavity mode Fock states)
is to exchange the last two components $|110\rangle$ and $|111\rangle$.
The implementation of this gate needs the same arrangement 
of aligned cavities crossed by Rydberg atoms
used for the C-NOT gate of Fig. $3$, 
except that now one has {\em three} cavity qubits  
(with the corresponding classical sources 
$S_{A}$, $S_{B}$ and $S_{C}$) instead of two. The  
auxiliary cavity $M$ is again needed for the atomic mirror scheme 
used to disentangle the atom.
The atom is initially prepared in state $|g\rangle$ and when it is in 
the first cavity $A$, is subject to a 
$\pi$ pulse between the dressed state $| {\cal V}^{(0)}_{+}(0) \rangle$ and 
$|g,0\rangle$. This pulse creates atom-cavity entanglement and the
state of the total system becomes
\begin{equation}
\Big[ \alpha_{1}|000\rangle   +\alpha_{2}|001\rangle +\alpha_{3}|010\rangle 
+\alpha_{4}|011\rangle \Big] \otimes | e \rangle +
\Big[ \alpha_{5}|100\rangle +\alpha_{6}|101\rangle +\alpha_{7}|110\rangle 
+\alpha_{8}|111\rangle \Big] \otimes | g \rangle.
\label{ok3}
\end{equation}
Then, when the atom reaches the second cavity $B$, it undergoes 
another $\pi$ pulse, at the new frequency $\omega_{2}$
corresponding to the transition between $|g,0\rangle$ and 
$| {\cal V}^{(2)}_{+}(0) 
\rangle$, so that the transformation
\begin{equation}
|0\rangle_{B}\otimes | g \rangle \rightarrow |2\rangle_{B}\otimes | e 
\rangle,
\label{ok4}
\end{equation}
is realized. This means temporarily leaving the 
logical subspace, even though this allows us to realize
a significant simplification of the scheme.
The state after this second step is therefore
\begin{equation}
\Big[ \alpha_{1}|000\rangle   +\alpha_{2}|001\rangle  +  \alpha_{3}|010\rangle 
+\alpha_{4}|011\rangle  + \alpha_{5}|120\rangle +
\alpha_{6}|121\rangle \Big] \otimes |e 
\rangle + \Big[ \alpha_{7}|110\rangle 
+\alpha_{8}|111\rangle \Big] \otimes | g \rangle.
\label{ok5}
\end{equation}
When the atom enters in $C$, the classical field $S_{C}$ is 
applied so to realize the C-NOT(atom $\rightarrow$ 
cavity $C$) of Sec.~III and the state of Eq.~(\ref{ok5}) becomes
\begin{equation}
\Big[ \alpha_{1}|000\rangle   +\alpha_{2}|001\rangle  +  \alpha_{3}|010\rangle 
+\alpha_{4}|011\rangle +  \alpha_{5}|120\rangle +\alpha_{6}|121\rangle \Big] \otimes |e 
\rangle + \Big[ \alpha_{7}|111\rangle 
+\alpha_{8}|110\rangle \Big] \otimes | g \rangle.
\label{ok6}
\end{equation}
At this point one has to
disentangle the atom from the three cavities and also to adjust
the state components in which the cavity $C$ contains two photons.
Both problems can be solved using again the
auxiliary cavity $M$ and a second atom crossing the apparatus in the 
opposite direction as in the ``atomic mirror'' configuration of 
Sec.~III. The cavity $M$ transfers the entanglement with the cavities
from the first to the second atom, which  
is not subject to any classical pulse in $C$. Then the second atom
enters $B$, where it undergoes a $\pi$ pulse at the frequency 
$\omega_{2}$, which simply inverts the transformation of
Eq.~(\ref{ok4}) (thanks to the fact that no $|0\rangle_{B}\otimes|g \rangle$
term is present), correcting in this way the terms of Eq. (\ref{ok6}) in which 
the second cavity contains two photons. 
As a consequence, the state of the atom-cavities system becomes
\begin{equation}
\Big[ \alpha_{1}|000\rangle   +\alpha_{2}|001\rangle  +  \alpha_{3}|010\rangle 
+\alpha_{4}|011\rangle \Big] \otimes |e \rangle +  \Big[\alpha_{5}|100\rangle
 +\alpha_{6}|101\rangle + \alpha_{7}|111\rangle 
+\alpha_{8}|110\rangle \Big] \otimes | g \rangle.
\label{ok7}
\end{equation}
Finally the atom enters $A$, where it is subjected to a 
$\pi$ pulse resonant with the transition $| {\cal V}^{(0)}_{+}(0) \rangle
\rightarrow  |g,0\rangle$, exchanging
$|0\rangle_{A}\otimes |g \rangle$ with $|0\rangle_{A}\otimes |e 
\rangle$, so that the second atom is disentangled from the cavities
and one gets the desired generic output of a Toffoli gate, i.e., 
\begin{equation}
\Big[ \alpha_{1}|000\rangle   +\alpha_{2}|001\rangle  +  \alpha_{3}|010\rangle 
+\alpha_{4}|011\rangle + \alpha_{5}|100\rangle
 +\alpha_{6}|101\rangle + \alpha_{7}|111\rangle 
+\alpha_{8}|110\rangle \Big] \otimes | g \rangle .
\label{ok8}
\end{equation} 
Notice that in this way we have implemented the Toffoli gate with 
two atoms only, as in the C-NOT gate of Sec.~III. Moreover 
this scheme can be easily extended to the case of $n \geq 4$ 
cavity qubits, for the implementation of the n-qubit generalization
of the Toffoli gate. We need only two atoms crossing the
aligned cavities in opposite directions also in this more general case. 
The pulse sequence
is similar to that discussed above: both atoms undergo 
a $\pi$ pulse resonant with the transition $| {\cal V}^{(0)}_{+}(0) \rangle
\rightarrow  |g,0\rangle$ in the first cavity,  
while in the following $n-2$ cavities they are submitted to a
$\pi$ pulse at the frequency $\omega_{2}$. In the last cavity, 
the target qubit, the first atom experiences a 
C-NOT(atom $\rightarrow$ 
cavity) while the second atom crosses it undisturbed.
In the scheme proposed here, the target qubit is necessarily the last 
cavity.
However, it is always possible 
to realize the n-qubit generalized Toffoli gate with the target 
qubit in a generic position of the string of cavities, 
by simply applying a two-qubit operation to the above scheme. 
In our scheme  
this means using four atoms at most, and in any case
this is much more convenient than realizing this generic
n-qubit gate from one and two qubits operation.

\subsection{Encoding and decoding in quantum error correction codes}

Other examples for which the present cavity QED scheme 
offers the possibility of an efficient implementation
of operations involving many qubits are the encoding and decoding
processes used in quantum error correction schemes \cite{Lafla,Knill}. 
Errors in quantum information processing are due to the interaction 
with uncontrolled degrees of freedom (the environment), yielding
an entanglement of the quantum state of the register with some 
environmental states. The main idea of quantum error correction is
to combat ``bad'' entanglement with ``good'' entanglement, that is,
protecting quantum information by storing it not 
in a single qubit but in an entangled state of $n$ qubits.
This is the encoding process; if the error rate is not too large,
it is possible to recover
the original quantum information by using a suitable decoding procedure,  
because the eventual data corruption
can be revealed by a measurement on the auxiliary qubits and 
information can be finally restored with single qubit operations.
The more general one qubit error (flip error, phase error, or a 
combination of the two) can always be corrected
using a five-qubit encoding and decoding procedure \cite{Lafla,Knill}.
However, if one considers one specific form of error only,
three-qubit encoding is sufficient for the 
implementation of quantum error correction codes.
For simplicity, let us consider this latter case. 

Let us assume that we want to protect a generic state
$\alpha | 0 \rangle_{A} + \beta | 1 \rangle_{A} $  
of the cavity $A$. For the encoding process one needs two other ancilla
qubits, cavity $B$ and $C$, and one has to realize the following
transformation into a 
maximally entangled, GHZ state \cite{ghz} of three cavities
\begin{equation}
 \Big[ \, \alpha | 0 \rangle_{A} + \beta | 1 \rangle_{A} \,
  \Big] \, \otimes | 0 \rangle_{B}
  \otimes | 0 \rangle_{C} 
  \rightarrow 
  \alpha | 0 0 0 \rangle + \beta | 1 1 1 \rangle .
 \label{encod}
 \end{equation}
This encoding process can be realized using a scheme
analogous to those discussed above for the C-NOT and Toffoli gates.
 Again, only two atoms are needed, with the
second one, crossing the aligned cavities in opposite direction, which
serves the purpose of disentangling the first atom, with the help
of the auxiliary cavity $M$ in the ``atomic mirror'' scheme described
in Sec.~III. 

The initial state of the system is
\begin{equation}
 \Big[ \, \alpha | 0 \rangle_{A} + \beta | 1 \rangle_{A} \,
  \Big] \, \otimes | 0 \rangle_{B}
  \otimes | 0 \rangle_{C} \otimes |e \rangle 
  \equiv
   \Big[ \, \alpha | 0 0 0 \rangle + \beta | 1 0 0 \rangle  \, 
  \Big] \,  | e \rangle .
 \label{sette2}
 \end{equation} 
When the atom enters $A$ it undergoes the 
C-NOT(cavity $A$ $\rightarrow$ atom) described by Eq.~(\ref{cinque});
when the atom arrives in $B$, the classical field $S_{B}$ is switched on in 
order to realize the C-NOT(atom $\rightarrow$ cavity $B$) transformation
(see Sec.~III) and the same C-NOT(atom $\rightarrow$ cavity $C$)
operation is applied when the atom is in $C$.
Using Eqs.~(\ref{cinque}) and (\ref{sei}), one can show that the
state of the total system becomes
  \begin{equation}
  \Big[  \, \alpha | 0 0 0 \rangle | e \rangle + \beta | 1 
  11 \rangle  | g \rangle \,  \Big].  
  	\label{otto}
  \end{equation} 
Except for the entanglement with the atom, the 
state of $A$,$B$ and $C$ is of the desired form, and therefore the 
situation is analogous to that of the C-NOT gate of Sec.~III (see 
Eq.~(\ref{nove1})). Atom disentanglement can be obtained
using again the atomic mirror scheme of Sec.~III, or eventually, the 
atomic detection scheme discussed in Sec.~IV, which is however
seriously limited by detector inefficiencies.
The logical transformations we have implemented in this section is 
schematically described in Fig.~6 (the dotted line box represents the 
atomic mirror)

Decoding is obtained
by repeating exactly the same procedure adopted for encoding the state
and assuming as initial condition for the qubits set the encoded state 
(\ref{encod}). 
It is straightforward to check that this amounts to realize the 
inverse transformation
  \begin{equation}
   \alpha | 0 0 0 \rangle + \beta | 1 
  11 \rangle  \rightarrow  \alpha | 0 0 0 \rangle + \beta | 1 
  0 0 \rangle = \Big[ \, \alpha | 0 \rangle_{A} + \beta | 1  \rangle_{A}
  \, \Big] \otimes | 0 \rangle_{B} \otimes | 0 \rangle_{C}.
  \label{dodici}
  \end{equation} 
 
It is also easy to see that the encoding scheme
described here can be extended for the controlled preparation of
maximally entangled states of $n$ cavities.  
One has to consider $n$ cavities (plus the auxiliary cavity $M$) and,
as in the $n=3$ case, one needs only two atoms crossing in opposite 
directions the $n+1$ aligned cavities. 
In analogy with the description above, in the first cavity the
two atoms undergo the C-NOT(cavity $A$ $\rightarrow$ atom) transformation
of Eq. (\ref{cinque}), while in all the others $n-1$ cavities 
the first atom undergoes to a C-NOT(atom $\rightarrow$ cavity) gate
and the second one crosses them with no classical field applied. 
In this way, thanks to the disentanglement action of the second atom,
the following maximally entangled state of $n$ cavities is prepared
\begin{equation}
 \Big[ \, \alpha | 0 \rangle_{1} + \beta | 1 \rangle_{1} \,
  \Big]  | 0 \rangle_{2}
  \ldots  | 0 \rangle_{n} 
  \rightarrow 
  \alpha | 0 0 \ldots 0 \rangle + \beta | 1 1 \ldots 1 \rangle .
 \label{mes}
 \end{equation} 

\section{Conclusions}

In this paper we have presented a scheme for implementing quantum logic 
operations within a cavity QED configuration. 
The quantum register is composed by a series
of high-Q cavities and information is encoded in the two lowest cavity
Fock states. Both the preparation and the detection of the quantum 
state of individual qubits, which is an essential ingredient in 
quantum algorithms, can be easily performed. In particular the 
detection of the two Fock states could be even performed in a quantum 
non-demolition way, as recently demonstrated \cite{haros}.
Both one-qubit and two-qubit operations can be performed sending
appropriately prepared atoms through the cavities.
An important advantage of the scheme is that it is particularly 
suitable for the direct implementation of some useful many-qubit quantum 
gates, as for example the Toffoli gate and its n-qubit generalization, 
or the encoding-decoding network of quantum error correction codes.
The scheme could be implemented in a generic cavity QED scheme, even if
here we have specialized to the case of microwave cavities and
circular Rydberg atoms, for which entanglement manipulation has been
already demonstrated \cite{qpg,haros}. 
For example, the same scheme could be adapted
to the optical frequency domain, by using high-Q optical cavities, as for
example the whispering gallery modes of silica microspheres \cite{wgm}, in
which one can have a miniaturization of the scheme and, therefore, a faster
gate operation. The scheme proposed here is in principle scalable, 
even if in practice its scalability
will be limited by various facts. In the microwave case one is limited 
by the spontaneous emission from the circular Rydberg levels ($\simeq 30$
ms) and by the fact that all the apparatus has to be cooled at cryogenic
temperatures to avoid the thermal radiation. In the optical case
cooling is no more needed and also the limitations due to the spontaneous
emission could be avoided in principle by using atomic
$\Lambda $ transitions
and adiabatic passage through a dark state \cite{adia}. However
in the microwave case, the proposed quantum gates
could be implemented using available technology, and therefore 
proof of principle demonstrations of quantum computation with, say, ten
qubits could be achievable. Instead, even if whispering gallery modes
in microspheres with $Q \simeq 10^9$ has been already realized \cite{wgm}, 
entanglement manipulation in this case 
have not been experimentally demonstrated yet.

Recently, a similar proposal of using the two lowest Fock states
of a high-Q cavity as logical qubit has been presented, involving 
an engineered network of defects in a photonic band-gap material \cite{pbg}.
This proposal is promising with respect to scalability since,
using atoms travelling along engineered waveguides in the photonic
band-gap material, spontaneous emission could be completely eliminated.
However, even if this proposal is promising in terms of the technological
realization, in this case, as in the silica microsphere case, 
entanglement manipulation has not yet been experimentally demonstrated.

From a general point of view 
the scheme proposed here is analogous to the linear ion trap scheme, 
except that
now the high-Q cavities play the role of the ions, and the atoms, and not the
collective center-of-mass motion, play the role of the quantum bus.
At first sight, it may seem unpratical to reverse the role of atoms and photons
as we have done here, since the common wisdom is that
atoms and ions are suitable for {\em storing} informations while
photons are best suited to {\em transfer} quantum information between 
different sites. However, the practical implementation of quantum algorithms
on linear ion traps is presently limited by the heating of the center-of
mass motion \cite{heat}. On the contrary, in the case of photonic qubits
discussed here, once
the limitations due to the spontaneous emission are eliminated (as in the
photonic band-gap case and in the microsphere case with dark state transitions),
the scheme is then only limited by the decoherence due to the finite $Q$ of the 
cavities, which could reach however values of the order of $10^{10}$, 
allowing therefore a sufficient number of gate operations.
Moreover, in view of the fact that photons are in any case the best tools
for quantum information transport, it may be nonetheless
useful to have schemes able
to process and temporarily store quantum information using photons.

\appendix

\section{A realization of the Deutsch algorithm}

As an example of a simple algorithm which can be implemented with the 
quantum gates presented before we consider the 
Deutsch algorithm \cite{Deut1} in the improved version of Ref. 
\cite{Deut2}. 

Consider a generic Boolean function $f(x)$ mapping $\{0,1\}$ into $\{0,1\}$. 
There are four different possibilities, the two constant functions
$ f_{1}(x) = 0$, $f_{2}(x) = 1$, and the two {\em balanced} functions
$f_{3}(x) = x$, 
$f_{4}(x) = 1-x$.
Using classical algorithms, the distinction between these 
two different classes of 
functions necessarily requires that {\em both} values  
of $f(x)$ have to be evaluated. 
On the contrary, the Deutsch quantum algorithm solves
the problem in just one step. We have to consider the two cavities
quantum circuit of Fig.~7. Initially both cavities 
 are in the vacuum state; then they are submitted to 
the Hadamard-phase transform of equation (\ref{1quattro}),
with phases equal to $0$ and $\pi$ 
respectively (step $a)$ of Fig.~7). 
As shown in Sec.~IV, this transformations can be 
implemented with a single atom prepared in the ground state $|g 
\rangle$ crossing both cavities.
At this point the system undergoes the transformation of step 
$b)$, namely the following ``$f$'' gate:
\begin{equation}
|x\rangle_{A}|y\rangle_{B} \rightarrow | x \rangle_{A} \left | \Big[ y + f_{i}(x) 
\Big]_{mod2} \right\rangle_{B},
\label{w3}
\end{equation}
where $f_{i}(x)$, $i=1,2,3,4$ are the four functions
defined above, and  $\Big[ y + f_{i}(x) \Big]_{mod2}$
means addition modulo $2$.
For $i=1$, this means that nothing happens 
to the system; on the 
contrary for $i=2$ $A$ remains unchanged but the second qubit in 
cavity $B$ undergoes a NOT transformation. 
Finally it is easy to 
verify that for $i=3$ and for $i=4$ the $f$ gate of Eq.~(\ref{w3})
is equivalent to the C-NOT gate
and to the C-NOT-INV gate respectively. In the preceding sections we have 
seen how to implement all these transformations.
At stage $c)$ we have then to implement another Hadamard 
transformation with zero phase on $A$ and measure the state
of this qubit. According to the logical network of Fig.~7, it is
possible to show that if the $f_{i}(x)$ of the $f$ gate
is a constant function, then the cavity $A$ must be found in 
$|0 \rangle_{A}$, otherwise, if
$f_{i}(x)$ is a balanced function, the cavity $A$ will be in $| 1 
\rangle_{A}$: in this way, one can establish if the function $f_{i}$
is constant or balanced using a single function evaluation.
The physical implementation of the Deutsch
problem in terms of the cavities is sketched in Fig.~8.
In the case of the constant function it is possible to only use two 
atoms, because for $i=2$ 
one atom is needed for the NOT transformation on the cavity 
$B$ of stage $b)$ and another atom is needed for the Hadamard
transform on the cavity $A$ at step $c)$.
For the balanced functions, the 
number of atoms is instead equal to four, because,
besides the two atoms for the implementation of the
Hadamard transformations, one has to use two atoms for the 
C-NOT (if $i=3$) or the C-NOT-INV (if $i=4$).

\begin{figure}
\centerline{\epsfig{figure=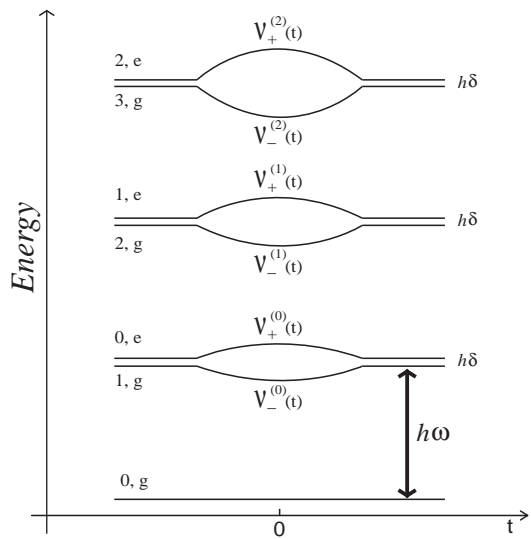,width=7cm}}
\vspace{0.2cm}
\caption{Energy level of the dressed states as a function of time.}
\label{figura1}
\end{figure}

\begin{figure}
\centerline{\epsfig{figure=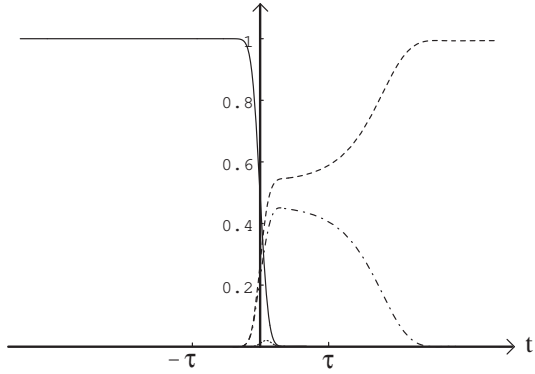,width=7cm}}
\vspace{0.2cm}
\caption{Time evolution of the population of the dressed states 
$|g,0\rangle$ (full line), $| {\cal V}^{(0)}_{-}(t) \rangle$ (dashed 
line), $| {\cal V}^{(0)}_{+}(t) \rangle$ (dotted-dashed line) and $| 
{\cal V}^{(1)}_{+}(t) \rangle$ (dotted line)
for the C-NOT(atom $\rightarrow$ cavity) gate with $\Omega_{0}=420$ kHz, 
$\delta= 0.18 \,\Omega_{0}$,
  $\tau \sim 100 \mu s$, $\Xi_{0}=141.5$ kHz, $\tau_{S}=19 \mu s$ and 
  initial condition $|\psi_{0}\rangle = | g , 0 \rangle$.}
\label{figura3}
\end{figure}

\begin{figure}
\centerline{\epsfig{figure=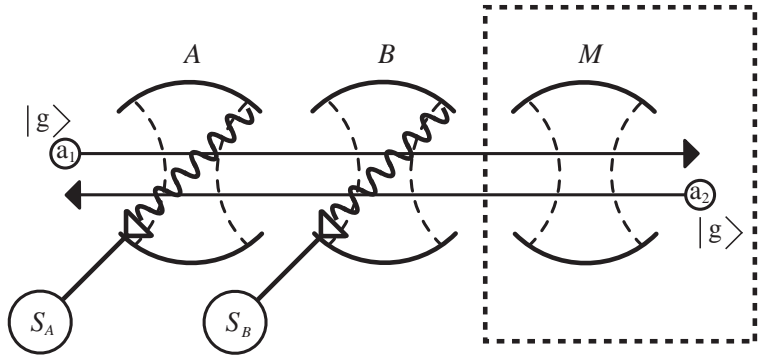,width=10cm}}
\vspace{0.2cm}
\caption{Schematical description of the C-NOT gate in which cavity $A$
is the control qubit and cavity $B$ the target qubit. $a_{1}$ and 
$a_{2}$ are the two atoms, and $M$ is the auxiliary cavity, 
transferring the entanglement with the cavities 
from the first to the second atom.}
\label{figura4}
\end{figure}

\begin{figure}
\centerline{\epsfig{figure=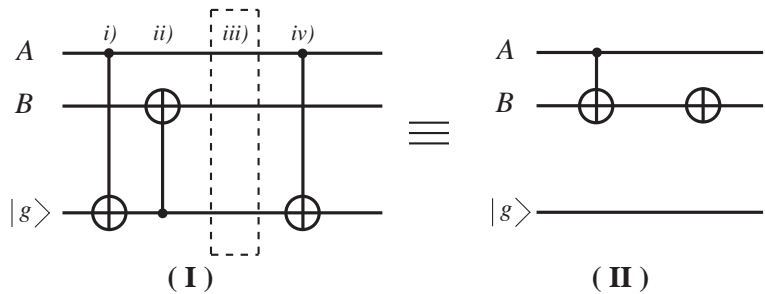,width=10cm}}
\vspace{0.2cm}
\caption{Logical scheme of C-NOT-INV gate of Fig.~3. The dashed box 
denotes the ``atomic mirror'' (see Sec.~III).}
\label{figura5}
\end{figure}

\begin{figure}
\centerline{\epsfig{figure=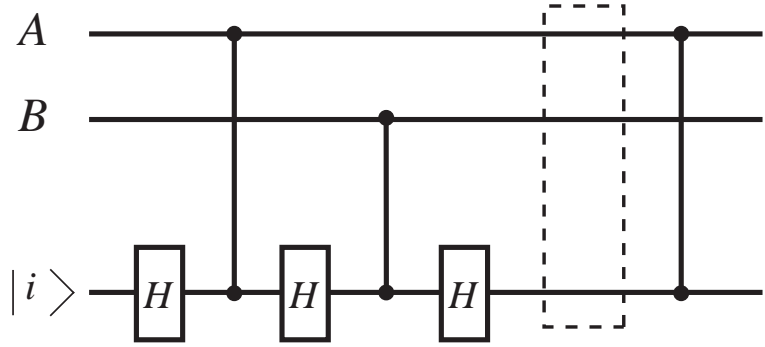,width=10cm}}
\vspace{0.2cm}
\caption{Logical scheme of the quantum phase gate (QPG) between two cavities.
The vertical lines denote the QPG, while $H$ denotes the Hadamard 
gates (i.e., $\pi/2$ pulses) applied on the atom. The dashed box 
denotes the ``atomic mirror'' (see Sec.~III).}
\label{figura5b}
\end{figure}

\begin{figure}
\centerline{\epsfig{figure=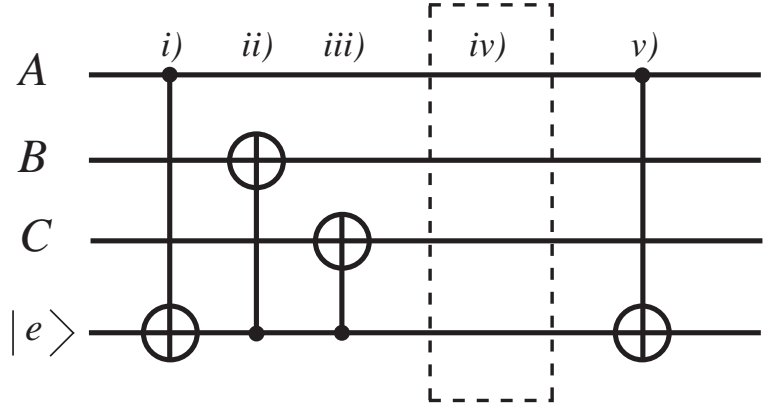,width=10cm}}
\vspace{0.2cm}
\caption{Logical scheme of the encoding network of 
Eq.~(\protect\ref{encod}); qubits $B$ 
and $C$ are initially prepared in the vacuum state, while $A$ is the 
qubit storing the initial information. The dashed box {\em iv)}
 represents the atomic mirror described in Section III.}
\label{GHZ3}
\end{figure}

\begin{figure}
\centerline{\epsfig{figure=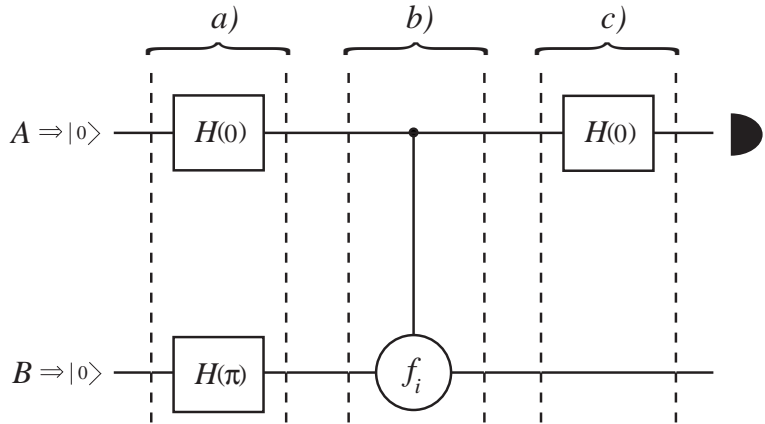,width=10cm}}
\vspace{0.2cm}
\caption{The Deutsch gate. The box with $H(\theta)$ performs a 
``Hadamard-phase'' transformation on the qubit with phase $\theta$ (see 
Eq.~(\protect\ref{1quattro}) ).}
\label{figura6}
\end{figure}

\begin{figure}
\centerline{\epsfig{figure=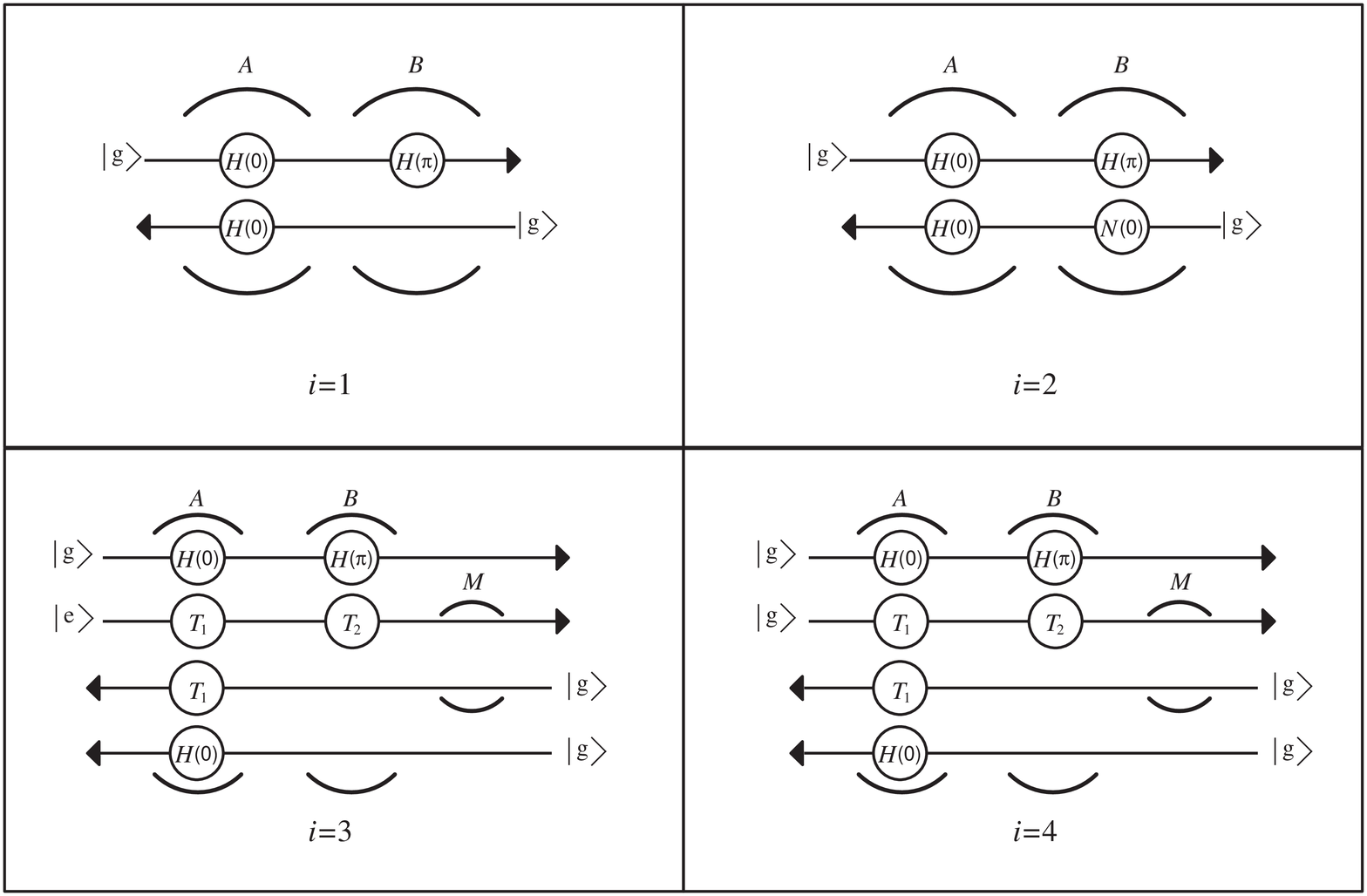,width=15cm}}
\vspace{0.2cm}
\caption{Physical implementation of the Deutsch gate: $N(\theta)$ and
$H(\theta^{\prime})$ are 
the one-qubit transformations of Eqs.~(\protect\ref{1tre}) and 
(\protect\ref{1quattro});  
$T_{1}$, $T_{2}$ are the C-NOT(atom $\rightarrow$ 
cavity) and C-NOT(cavity $\rightarrow$ 
atom) respectively. The index $i$ correspond to the four different 
functions $f_{i}(x)$.}
\label{figura7}
\end{figure}

\end{document}